\documentclass[prb, twocolumn, nofootinbib, citeautoscript, 10pt, longbibliography, notitlepage]{revtex4-2}

\usepackage{graphicx}
\usepackage{dcolumn}
\usepackage{bm}
\usepackage{amsmath}
\usepackage{tabularx,graphicx}
\usepackage{epstopdf}
\usepackage{latexsym}
\usepackage{amssymb}
\usepackage{color, colortbl}
\usepackage{psfrag}
\usepackage{bbm}
\usepackage{titlesec}
\usepackage{dsfont}
\usepackage{feynmp}
\usepackage{slashed}
\usepackage{multirow}
\usepackage{physics}
\usepackage[tight]{subfigure}

\usepackage[papersize={8.5in,11in}]{geometry}

\usepackage{color}
\definecolor{darkblue}{rgb}{0.,0.,0.4}
\definecolor{darkred}{rgb}{0.5,0.,0.}
\definecolor{BlueViolet}{RGB}{138,43,226}
\definecolor{SkyBlue}{RGB}{30,144,255}
\definecolor{DarkGreen}{RGB}{0,100,0}
\usepackage[pdftex,colorlinks=true,linkcolor=darkblue,citecolor=blue,urlcolor=darkred]{hyperref}

\renewcommand{\epsilon}{\varepsilon}

\def \nn{\nonumber \\}

\begin{document}

\title{Characterizing topological pumping of charges in exactly solvable Rice-Mele chains of the non-Hermitian variety}

\author{Ipsita Mandal}
\email{ipsita.mandal@snu.edu.in}

\affiliation{Department of Physics, Shiv Nadar Institution of Eminence (SNIoE), Gautam Buddha Nagar, Uttar Pradesh 201314, India}

\begin{abstract}
We address the nature of the Thouless charge-pumping for a non-Hermitian (NH) generalization of the one-dimensional (1d) Rice-Mele model, considering a variety which allows closed-form analytical solutions for the eigensystems. The two-band system is subjected to an ``adiabatic'' time-periodic drive, which effectively gives us a closed two-dimensional (2d) manifold in the momentum-frequency space, mimicking a 2d Brillouin zone (BZ) for periodic boundary conditions. For open boundary conditions, we formulate the non-Bloch generalized Brillouin zone (GBZ). All these allow us to compute the Chern numbers of the 2d manifolds emerging for various parameter values, using the BZ or GBZ, which are well-defined integers in the gapped regions of the spectra. If Thouless pumping still holds for NH situations, the expectation values of the net displacement for the particles occupying a given band within a cycle, calculated using the right and left eigenvectors of a biorthogonal basis for the complex Hamiltonians, would coincide with the Chern numbers. However, we find that there are deviations from the expected quantized values whenever the spectra exhibit strongly-fluctuating magnitudes of the imaginary parts of the (generically) complex eigenvalues. This implies that we need to address the question of how to meaningfully define time-evolution, as well as adiabaticity, in the context of NH systems.
\end{abstract}

\maketitle

\tableofcontents

\section{Introduction}


The concept of topological pumping of charges, predicted by Thouless \cite{thouless1,thouless2}, refers to the periodic pumping of quasiparticles in a one-dimensional (1d) lattice during an adiabatic variation of the parameters of the system. The number of particles (or, charges, when we are considering itinerant electronic excitaions) pumped per cycle turns out to be an integer, reflecting the Chern number\footnote{We would like to point out that the Chern number can be defined for a 2d system only for a gapped spectrum.} of the $(1+1)$-dimensional (2d) manifold constructed out of the 1d Brillouin zone (BZ) and one period of the time-lapse. This quantized nature of the net transported charge is a quantum effect (i.e., absent in a classical theory), purely determined by the topology of the underlying manifold, which makes it robust against perturbations that do not cause a topological phase transition.
In this paper, we consider the charge pumping for a non-Hermitian (NH) generalization of such 1d \textit{Hamiltonians}, which naturally captures effects such as dissipation and coupling to the environment. We will especially investigate what happens for open boundary conditions (OBCs). This is because, for NH systems, the spectra resulting from OBCs are fundamentally different from those resulting from periodic boundary conditions (PBCs) and the open boundary
condition (OBC) spectra are completely different \cite{emil_review, flore-elisabet, flore-elisabet0, maria_emil, ips-bp, kang-emil, ips-yao-lee}. On the other hand, for Hermitian systems, apart from
the possibility of extra edge modes (compared to the case of the PBCs), the remaining eigenstates and eigenenergies comprise the bulk and are insensitive to the specific boundary conditions in the thermodynamic limit. Charge pumping in the bulk under the OBCs does not bring out any surprising behaviour, embodied in the fundamental principle of the bulk-boundary correspondence \cite{schnyder_review, ref_demand, ips-sudip, ips-ep-epl, ips-tewari}.  

We will demonstrate the NH charge pumping for an NH generalization of the 1d Su-Schrieffer-Heeger (SSH) model \cite{ssh0, ssh, yaowang,song_yao_ssh, han_ssh,  ssh_lee, yinjiangliluchen, maria_emil}, augmented by a staggered onsite potential at the sublattice sites (labelled as $A$ and $B$). The resulting 1d chain is known as the Rice-Mele model \cite{ricemele} with NH hoppings. While the original SSH model \cite{ssh0, ssh} is a bipartite chain, modelled to describe solitons in a polyacetylene chain which harbours topologically distinct phases, the original Rice-Mele model represents 1d semiconductors with broken inversion symmetry. Eventually, these theoretical models have been realized experimentally in quantum systems of widely distinct origins, such as cold-atoms in optical lattices \cite{ssh_expt}, a monolayer of chlorine atoms on a copper surface \cite{ssh_expt2}, and attractive ultracold fermions in a shaken optical lattice \cite{ricemele_optical}. Supplemented by the fact that such systems allow closed-form analytical solutions for the eigensystems, as long as we restrict ourselves to models with nearest-neighbour hopping terms only, they have emerged as the most popular prototypes to study topological properties in 1d systems, both in the Hermitian and NH contexts \cite{ssh_lee, yaowang, flore-elisabet, flore-elisabet0, maria_emil, ips-bp}. The possibility to formulate exact solutions of the OBC wavefunctions \cite{flore-elisabet, flore-elisabet0, maria_emil, ips-bp}, thus, has dictated our choice of using an NH Rice-Mele chain. 


\section{Background and formalism}
\label{secmethods}

In this section, we review the concepts and formalism of implementing biorthogonal bases and constructing generalized Brillouin zones, which are the tools we will use to characterize NH charge pumping. We also spell out the explicit form of the Hamiltonian that we are going to analyze using these tools.

\subsection{Biorthogonal basis and non-Hermitian Chern number}
\label{secbi}

We consider a periodically driven NH Hamiltonian defined on a 1d lattice. Due to the discrete lattice translational symmetry, it is convenient to express the time-dependent Hamiltonian in the momentum space as $  H (k,t)$, where $t$ and $k$ represent the time and the momentum variables, respectively. This represents the PBC case, because the Fourier transformation to the momentum space is allowed in the first place only for a periodic system. Due to the cyclic nature of the drive, the system is periodic in the time-direction as well. We define the time period as $T$.

First we review the way Chern numbers are defined for NH systems using biorthogonal bases, applicable for the diagonalization of generic complex matrices \cite{brody}. The discussion closely follows the details explained in Ref.~\cite{zhenming}.
Let us denote the instantaneous left and right eigenvectors, corresponding to the n$^{\rm th}$ band, as $|u^L_{n} (k, t)\rangle$ and $|u^R_{n} (k, t)\rangle $, respectively. We adopt a biorthogonal basis such that these eigenvectors satisfy the condition $\langle u^L_{n} (k, t)|u^R_{n} (k, t)\rangle = 1$. Hence, for the n$^{\rm th}$ band, the Chern number can be defined as
\begin{align}
\label{eq:pbc_Chern}
C_n &= \int^T_0 dt \int_{BZ}
\frac{dk}{2\,\pi} \, \Omega^{n}_{kt}
\nn & 
= - \,i\int^T_0 dt\int_{BZ}
\frac{dk}{2 \,\pi} \,
\Big [ \left \langle \partial_k u^L_{n} (k, t)|
\partial_t u^R_{n} (k, t) \right \rangle
\nn & \hspace{ 3.5 cm } 
- \left \langle \partial_tu^L_{n} (k, t)|\partial_k u^R_{n} (k, t)
\right \rangle \Big ] ,
\end{align}
where $\Omega^{n}_{k t}$ denotes the biorthogonal Berry curvature of the effective $(1+1)$-dimensional closed manifold.
We call it a Chern number because it does take a quantized integer value when the biorthogonal basis is used.

Under a periodic drive, the particle-pumping per cycle can be found by calculating the net biorthogonal displacement (BOD), 
\begin{align}
\Delta\bar{x}^{LR}_n \equiv \int^T_0 dt
\int_{BZ}\frac{dk}{2\,\pi} \,v_{n} (k,t)\,,
\end{align}
where
\begin{align}
\label{eq:pbc_velo}
v_n (k, t) & \equiv 
 \left \langle \tilde{u}^L_{n}(k, t)|
\left [\partial_k\hat{H}(k,t) \right ] |\tilde{u}^R_{n}(k, t) \right \rangle,
\nn \partial_t\bar{x}^{LR}_n(t) &= 
\int_{BZ}\frac{dk}{2\,\pi} \,v_{n} (k,t)\,.
\end{align}
Here, $\tilde{u}^R_{n}(k, t)$ [$\tilde{u}^L_{n}(k,t)$] denotes the time evolved right [left] eigenstate under the action of ${H}(k,t)$ [$ {H}^\dagger(k,t)$]. Under the assumption of adiabatic evolution, using the leading-order correction term of the time-dependent perturbation theory, once can identify $v_n (k,t)$ as the $kt$-component of the Berry curvature $\Omega^{n}_{kt}$ plus the the group velocity, and $\Delta\bar{x}^{LR}_n$ as the biorthogonal Chern number $C_n$, analogous to the Hermitian cases. For a detailed derivation of this equivalence of Eqs.~\eqref{eq:pbc_Chern} and \eqref{eq:pbc_velo}, we refer the reader to Ref.~\cite{zhenming}.
It is important to discuss here the implications of applying the Thouless pumping to NH systems, because now the eigenvalues are in general complex. The consequence of the nonzero imaginary components of the spectrum can potentially lead to
significant departures from the condition of an adiabatic evolution, especially if these parts are large enough during a cycle. This question has been addressed to some extent in Ref.~\cite{muga}, and reexamined for specific models in Ref.~\cite{zhenming}. Here, we will show how this aspect affects the final characteristics of quantization (or, lack thereof) using the exact eigenvalues in the model under investigation.

\begin{figure*}[t]
\subfigure[]{\includegraphics[width=0.4 \linewidth]{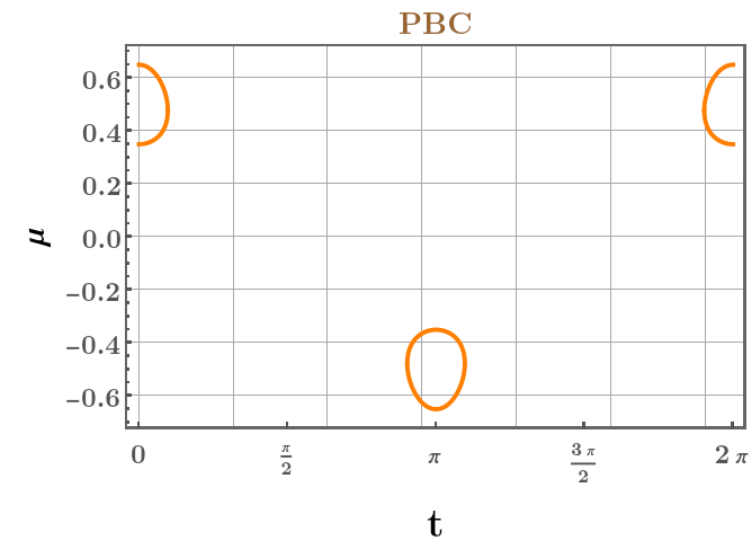}} 
\hspace{ 1 cm} \subfigure[]{\includegraphics[width=0.4 \linewidth]{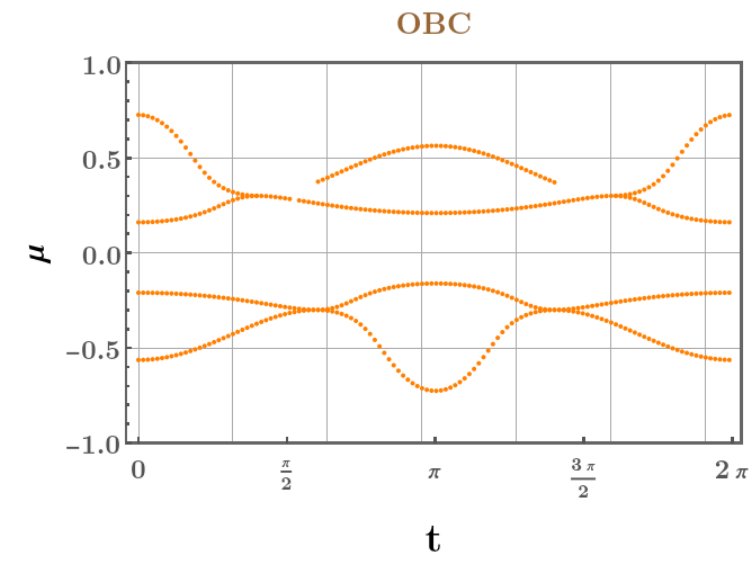}}
\subfigure[]{\includegraphics[width=0.38 \linewidth]{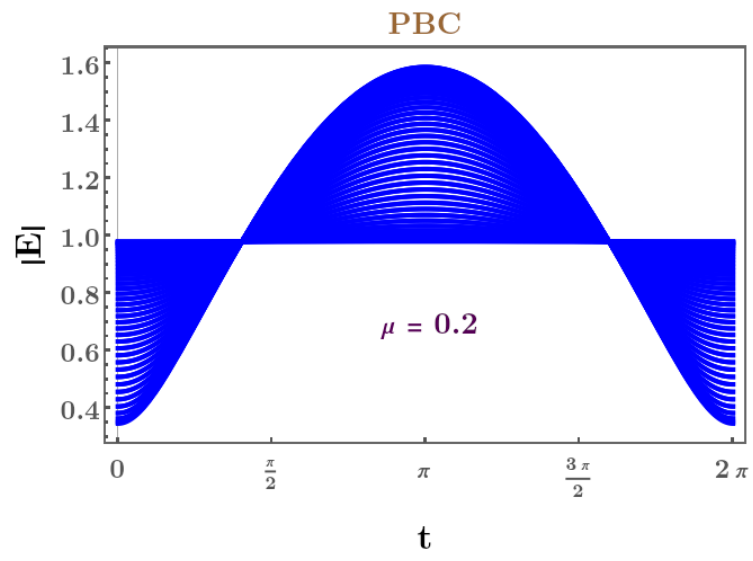}} 
\hspace{ 1 cm} \subfigure[]{\includegraphics[width=0.38 \linewidth]{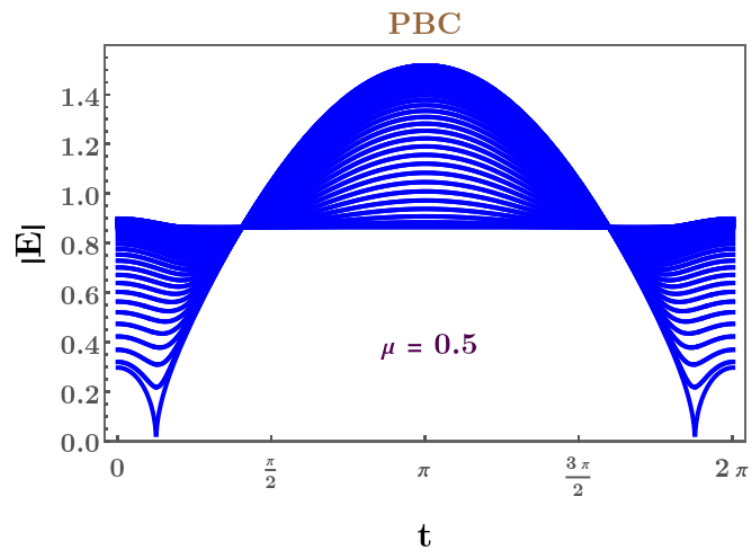}}
\subfigure[]{\includegraphics[width=0.38 \linewidth]{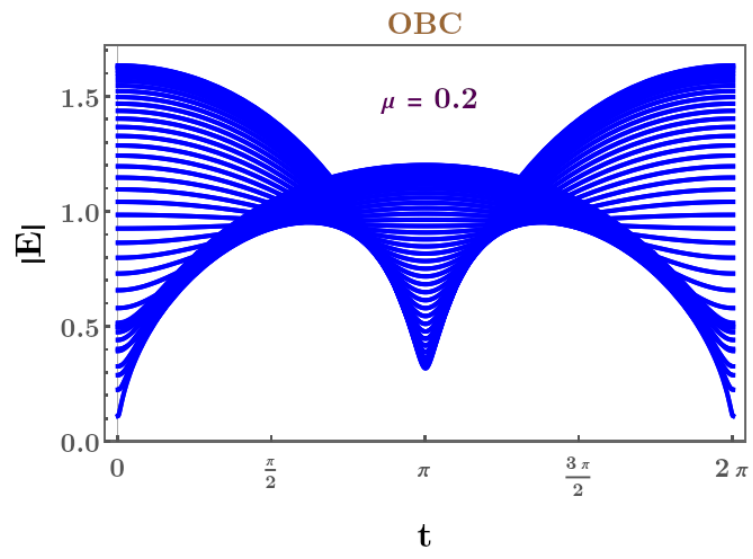}} 
\hspace{ 1 cm} \subfigure[]{\includegraphics[width=0.38  \linewidth]{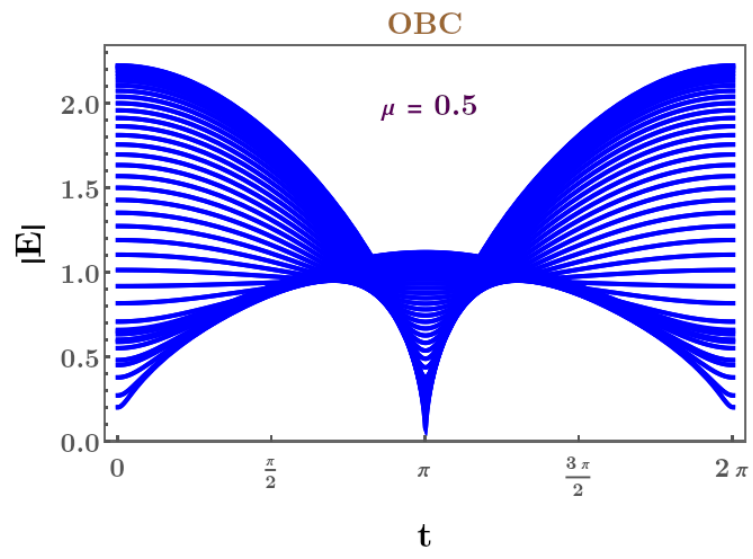}}
\caption{\label{figgap}For $\gamma =0.3 $, the uppermost panel shows the values of $\mu $ during one cycle of the drive, when the spectrum becomes gapless. Subfigures (a) and (b) represent the PBC and the OBC cases, respectively. The middle panel illustrates the spectra for two representative $\mu$-values, when the system with PBC is (c) gapped and (d) gapless (showing EPs). Likewise, the lowermost panel illustrates the spectra for two representative $\mu$-values, when the system with OBC is (e) gapped and (f) gapless.
}
\end{figure*}

The momentum-space representation of the PBC Hamiltonians, that we consider in this paper, gives rise to two-band systems of the form 
\begin{align}
\label{h_gen}
& H_{\rm PBC}(k,t) =  \mathbf{d} (k,t)  \cdot \boldsymbol \sigma \,,
\nn 
& \mathbf{d} (k, t) = \left \lbrace d_1 (k, t) , d_2 (k, t) , 
d_3 (k, t) \right \rbrace, \nn
& \boldsymbol \sigma  = 
\left \lbrace \sigma_1, \sigma_2, \sigma_3
\right \rbrace.
\end{align}
The right eigenvalues ($= \pm E_\textrm{PBC} $) and the corresponding right eigenvectors are given by
\begin{align}
& E_\textrm{PBC} (k,t)= 
\sqrt{d_1^2 (k, t) + d_2^2 (k, t) + d_3^2 (k, t) } \text{ and}
\nn & \psi^R_\pm (k, t) =\Big [ d_3 (k, t) \pm E_\textrm{PBC}
\quad  d_1(k, t) + i\,d_2 (k, t) \Big  ]^T ,
\end{align}
respectively.

For studying an NH generalization of the Rice-Mele chain \cite{ricemele, flore_2d_chern}, we consider the tight-binding model
obtained by setting \cite{yaowang, song_yao_ssh,  han_ssh}
\begin{align}
\label{h_ssh}
d_1 = t_1 +  t_2   \cos k \,,\quad  
d_2 =  t_2  \sin k + i\,\gamma \,,\quad d_3 = \Delta \sin t \,,
\end{align}
where
\begin{align}
t_1 = \mu \,, \quad t_2 = \mu - \cos t\,,
\quad \Delta = 1\,.
\end{align}
This leads to
\begin{align}
\label{eqmodel1}
& H_{\rm PBC}(k,t) \nn & = 
  \begin {bmatrix} 
  \sin  t & 
  \mu + \gamma + e^{-i \, k} \, (\mu - \cos  t) \\
\mu  - \gamma +  e^{i \, k}  \,(\mu - \cos  t) 
 & - \sin  t
 \end{bmatrix} .    
\end{align}
By definition, the SSH version corresponds to $\Delta = 0 $, because the staggered onsite potential is absent.
We note that, for any matrix $ \mathcal M $ of the form $\mathbf{d} \cdot \boldsymbol \sigma $, the eigenvalues always appear in pairs of $ \pm\, E$. For $ d_3 = 0 $, there is an explicit chiral symmetry, expressed as $\sigma_3 \, \mathcal M \, \sigma_3 = - \,\mathcal M $. Hence, at $t=0$, $H_{\rm PBC}(k,t) $ has a chiral symmetry, leading to the notion of chiral quasiparticles with chiral charge $+1$ or $-1$, depending on whether they occupy the $+E$ or the $-E$ band.
The corresponding lattice Hamiltonian in the real space is given by
\begin{align}
\label{ssh_real}
H &= \sum \limits_\ell \Big [ \left( t_1+ \gamma \right) 
c^\dagger_{\ell ,A} \,c_{\ell ,B}
+ \left( t_1 -  \gamma  \right) 
c^\dagger_{\ell ,B} \,c_{\ell ,A}
\nn & \hspace{ 1.2 cm }
+ t_2  \left( c^\dagger_{ \ell+1,A} \,c_{\ell, B} 
+ c^\dagger_{ \ell, B} \,c_{ \ell+1,A} \right) 
\nn & \hspace{ 1.2 cm }
+ \Delta \sin t \left(   c^\dagger_{\ell , A} \,c_{\ell , A} 
- c^\dagger_{\ell, B} \,c_{\ell, B}  \right)
\Big ] \,,
\end{align}
where $A$ and $B$ denote the two distinct sublattice sites in the bipartite lattice, and $\ell $ labels the lattice site.

\begin{figure*}[t]
\subfigure[]{\includegraphics[width=0.44\linewidth]{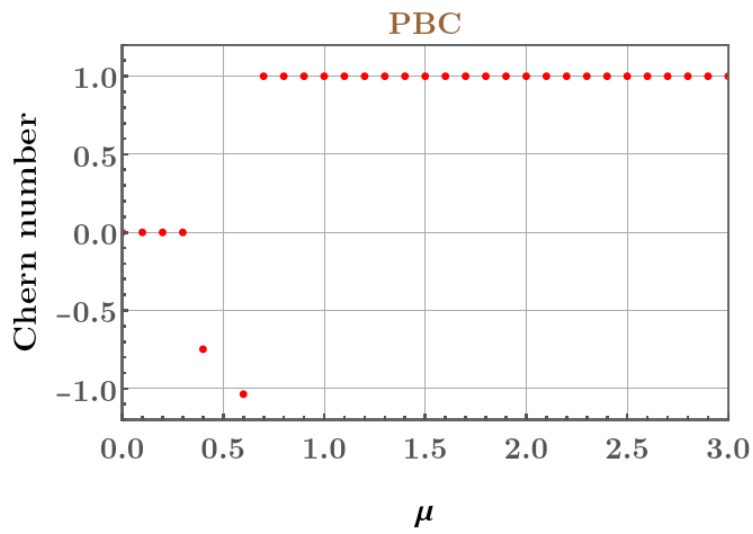}} \hspace{ 1 cm}
\subfigure[]{\includegraphics[width=0.43 \linewidth]{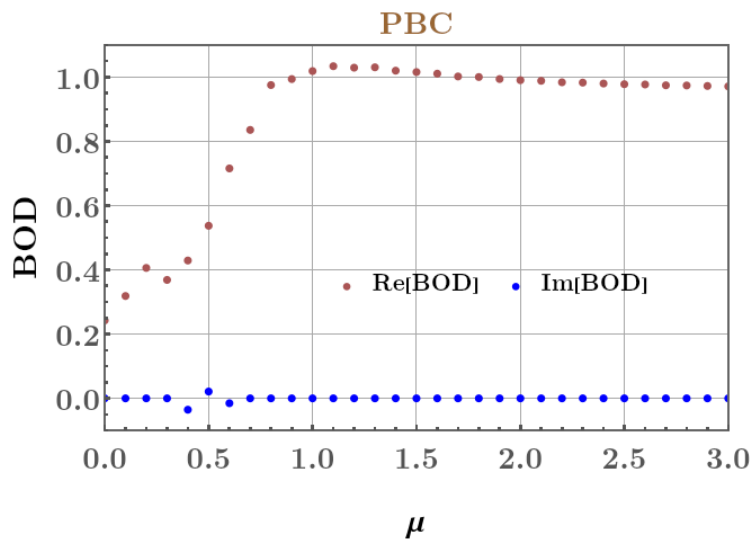}} 
\subfigure[]{\includegraphics[width=0.44 \linewidth]{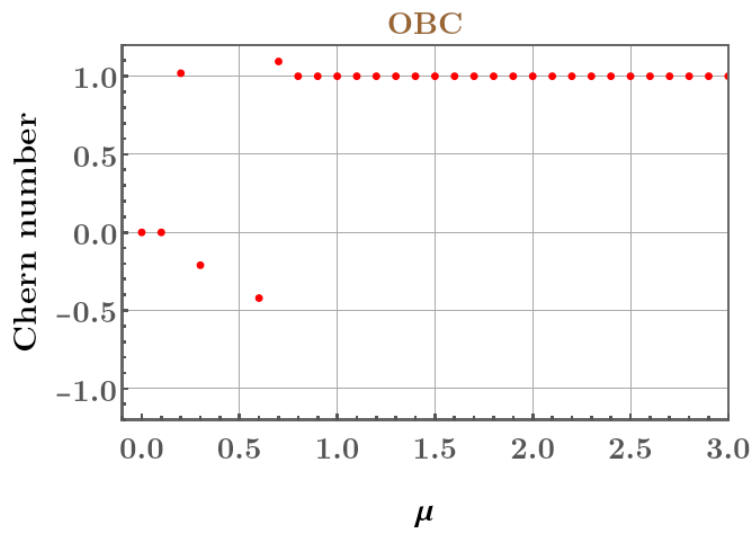}}
\hspace{ 1 cm} \subfigure[]{\includegraphics[width=0.44 \linewidth]{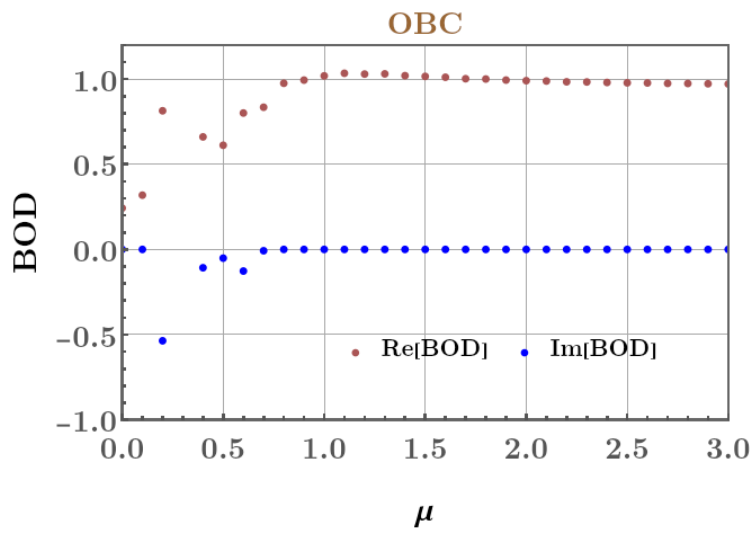}}
\caption{\label{figmod1} Chern number and biorthogonal displacement (BOD) for $\gamma =0.3 $. While subfigures (a) and (b) represent the PBC case [cf. Eq.~\eqref{eqmodel1}], subfigures (c) and (d) capture the OBC case [cf. Eq.~\eqref{eqmodel1obc}]. The Chern numbers show robust quantized plateaus in the gapped regions of the eigenvalue spectra, as one can confirm by comparing the $\mu$-axis values with Fig.~\ref{figgap}.
}
\end{figure*}

\subsection{Generalized Brillouin zone for OBCs}

For NH systems with open boundaries, the topological properties can be extracted by generalizing to the so-called non-Bloch band theory \cite{yaowang, yokomizo, ips-bp}. This involves using the concept of the generalized Brillouin zone (GBZ), denoted by the closed contour $ {\mathcal{C}}_{\text{GBZ}}$, by defining $\beta = e^{i\,k}$, where $k $ represents the momentum component for the direction along which OBC is imposed and is generically complex. The GBZ is determined by the following procedure: A generalized ``Bloch'' Hamiltonian $H(\beta)$ is obtained for the open system by substituting $k$ by $-i\,\ln \beta $ in the momentum-space Hamiltonian $H(k)$ for the corresponding periodic system. Therefore, it takes the form of
\begin{align}
\lbrace H(k), \, k \in \mathds{R} \rbrace \rightarrow 
\lbrace H(\beta=e^{i\,k}), \, k \in \mathds{C} \rbrace \,.
\end{align}
The eigenvalue equation $ \left [\beta^p \det[H(\beta) -E]  \right ]=0 $ gives a polynomial equation for $\beta$, where $p$ is the order of the pole of of $\det H$. If the degree of this equation is $2 M$, we need to arrange the roots $\beta_\varsigma $ (where $ \varsigma \in [1,\cdots, 2M]$) in the order
$|\beta_1| \leq |\beta_2|\leq \cdots \leq  |\beta_{2M-1}| \leq |\beta_{2M}|$. From this list, the curve for the GBZ is obtained from the trajectories of $\beta_{M}$ and $\beta_{M+1}$ under the condition $ | \beta_{M} | = |\beta_{M+1} | $. The energy spectrum $E$ of the open system is also obtained from these admissible values of the roots.

Defining $H(\beta, t) = \mathbf{d} (\beta, t) \cdot \boldsymbol{\sigma}$, the characteristic polynomial for $\beta $ takes the form of
\begin{align}
\label{eq_beta}
\beta^p  \, E^2_\textrm{PBC} ( \beta, t )
= \beta^p E^2  (k, t)\,.
\end{align}
Since the admissible roots have the same magnitude, they can be written as $\beta$ and $\beta\, e^{i\,\phi}$ [with $\phi \in [0,2\pi)$]. Substituting these two roots in the above equation, and subtracting the resulting equations from each other, we get
\begin{align}
\label{eq_beta2}
\beta^p\left [
 E^2_\textrm{PBC} (\beta, t)
 - E^2_\textrm{PBC} (\beta\, e^{i\,\phi}, t)
\right ]=  0 \,.
\end{align}
Here, the $E$-dependence has dropped out, and the equation is easier to solve. Hence, for $\phi$ taking values on the unit circle, we determine the roots $\beta_j$ (with $j \in [1,2M]$) for a given value of $\phi$, and pick out the admissible solutions, which satisfy $ |\beta_{M} | = |\beta_{M+1} |$, after ordering them in the ascending order according to their absolute values.

\begin{figure*}[t]
\subfigure[]{\includegraphics[width=0.44\linewidth]{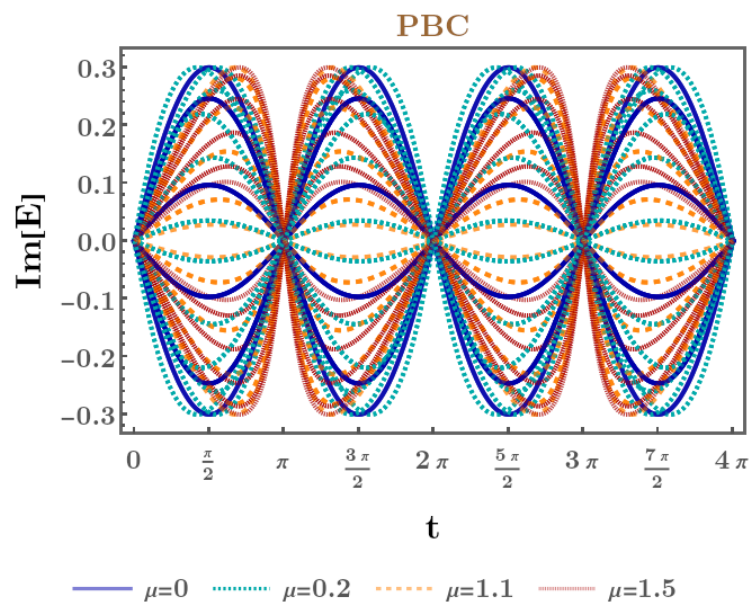}} \hspace{ 1 cm}
\subfigure[]{\includegraphics[width=0.43 \linewidth]{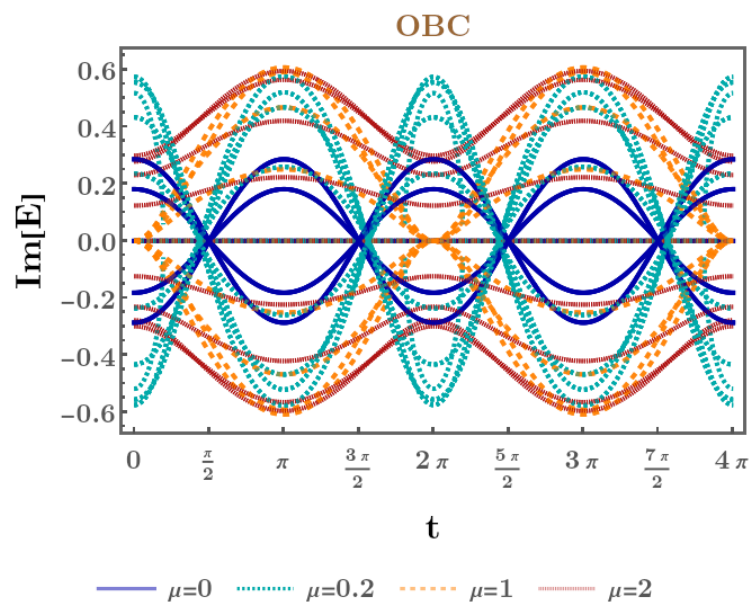}} 
\caption{\label{figim} Imaginary parts of the spectra for (a) PBC and (b) OBC, with $\gamma =0.3 $. The plotlegends in each subfigure show the values of $\mu$ that the curves represent.
}
\end{figure*}

For the values shown in Eq.~\eqref{h_ssh}, we find that Eq.~\eqref{eq_beta2} is quadratic in  $\beta$, which can 
be solved easily in closed forms. The allowed solutions are the ones when the two roots are equal in magnitude. In fact, we find that the solutions have $|\beta| = {\Gamma}$, with
\begin{align}
\Gamma =\sqrt{   \frac{| t_1 -\gamma| } {|  t_1 + \gamma|} } \,.
\end{align}
Since $\beta$ has a constant magnitude, the GBZ is a circle with radius $\Gamma$. Hence, with the simple substitution involving 
\begin{align} 
k  = \theta - i\ln \Gamma\,, \text{ with } \theta \in [0, 2\pi) \,,
\end{align} 
we define the effective Hamiltonian
\begin{align}
\label{eqhobc}
H_{\rm OBC} = H(\beta = \Gamma \,e^{i\,\theta})
\end{align}
for the OBC case~\cite{flore-elisabet0, flore-elisabet, ips-bp}. The resulting system is defined in a momentum space representative of the GBZ. Consequently, the eigenvalue spectrum as well as the eigenvectors for the open system can be derived easily as closed-form analytical expressions, using this prescription.

For the NH Rice-Mele model defined via Eqs.~\eqref{eqmodel1} and \eqref{ssh_real}, the mapping shown in Eq.~\eqref{eqhobc} gives us the effective Hamiltonian
\begin{align}
\label{eqmodel1obc}
H_{\rm OBC}( \theta,t) = 
  \begin {bmatrix} 
  \sin  t & 
  \mu + \gamma + 
 \frac { \mu \, - \, \cos  t } {\Gamma  \, e^{i \,\theta}} \\
\mu  - \gamma + \Gamma  \, e^{i \,\theta} \, (\mu - \cos  t)
 & - \sin  t
 \end{bmatrix} .    
\end{align}

\section{Results}
\label{secresult}

In this section, using the PBC and OBC Hamiltonians shown in Eqs.~\eqref{eqmodel1} and \eqref{eqmodel1obc}, we explicitly demonstrate the results for NH charge pumping.

\subsection{Charge pumping for PBC}

First we calculate the left and right eigenstates at $t=0$ by using the PBC Hamiltonian shown in Eq.~\eqref{eqmodel1}. This allows us to compute the Chern number after time-evolving these eigenstates using $H_{\rm PBC}( k,t)$.

We need to identify the regions of parameter space where the spectra are gapped, because the Chern number is ill-defined as soon as the spectrum becomes gapless. In the context of NH systems, the gaplessness often involve singular degeneracies known as the Exceptional Points (EPs) \cite{emil_review, ips-emil, kang-emil, ips-kang}. Let us choose to analyse our model system for $\gamma =0.3 $. Fig.~\ref{figgap}(a) shows the range of values of $\mu$ for which the spectrum becomes gapless.
We compute the Chern number and the BOD numerically, the results of which are shown in Figs.~\ref{figmod1}(a) and~\ref{figmod1}(b), respectively. As expected, the Chern number settles into well-defined quantized values in the gapped spectral regions, compatible with the information contained in Fig.~\ref{figgap}(a). The nature of the PBC spectra, when EPs are absent [present] is also illustrated in Fig.~\ref{figgap}(c) [Fig.~\ref{figgap}(d)].

As a consequence of complex eigenvalues, in line with the discussion in Sec.~\ref{secbi}, the quantization of Thouless pumping may fail, depending on the magnitude of the imaginary parts. These imaginary parts, for some representative $\mu$-values, have been shown in Fig.~\ref{figim}(a). Comparing with Fig.~\ref{figmod1}(b), we find that the fluctuations of the magnitude of the imaginary parts are very high where the BOD deviates from the expected quantized values.

\subsection{Charge pumping for OBC}

For the OBC case, we carry out the same steps as in for the PBC case, except that, here, we need to use the mapped Hamiltonian $H_{\rm OBC}( \theta,t)$ [cf. Eq.~\eqref{eqmodel1obc}]. Again, we choose the representative value of $\gamma = 0.3 $. The corresponding ranges of $\mu $-values giving rise to gapless spectra are shown in Fig.~\ref{figgap}(b).
Examples of the gapped and gapless spectra are demonstrated in Figs.~\ref{figgap}(e) and (f), respectively. 
In the gapped-eigenvalue regimes, we get plateaus for the Chern number curve shown in Fig.~\ref{figmod1}(c). The numerically-computed BOD is also illustrated in Fig.~\ref{figmod1}(d).

The imaginary parts of the spectra, for some representative $\mu$-values, have been shown in Fig.~\ref{figim}(b). Comparing with Fig.~\ref{figmod1}(d), we find that the fluctuations of of the magnitude of the imaginary parts are very high where the 
BOD deviates from the expected quantized values.

\section{Summary and outlook}

In this paper, we have addressed the observation of NH pumping using a periodically driven Rice-Mele model, generalized to have nonreciprocal hoppings. The system, modelled on a bipartite lattice, has only nearest-neighbour-hopping terms, which makes it possible to have closed-form analytical solutions for an open chain. This special feature allows us to determine the exact forms of the right and left eigenvectors for the OBC case, using which the Chern number can be computed in the gapped regions of the spectrum. For 1d Rice-Mele chains with cyclic time-dependence of various parameters, the charge-pumping question has been earlier addressed in Ref.~\cite{zhenming}. However, there, the authors have demonstrated the NH puming for the PBC and OBC cases using Wannier states, which necessitates the introduction of some approximations in using the eigenstates for the OBC. We have overcome this issue by using the exact solutions available for a class of 1d NH models without longer-ranged hoppings \cite{flore-elisabet, flore-elisabet0, maria_emil, ips-bp}.

From our results, we have found that the BOD takes a quantized real value (in the gapped parts of the spectrum) where the imaginary parts of the energy eigenvalues do not fluctuate too much within the span of a cycle. However, how to put a mathematical bound on this is an open question, which we have not tried to answer in this paper. This needs to be addressed by asking the more fundamental question, with respect to NH systems, regarding how to meaningfully define time-evolution which would capture the correct physical behaviour. We would like to emphasize that, unlike what is claimed in Ref.~\cite{zhenming}, quantized charge-pumping is not guaranteed on using a biorthogonal basis, even on using exact eigenstates. 

\section*{Acknowledgments}
We thank Emil J. Bergholtz for suggesting the problem. We are also grateful to Zhenming Zhang and Atanu Maity for useful discussions. 

\bibliography{ref}

\begin{thebibliography}{32}%
\makeatletter
\providecommand \@ifxundefined [1]{%
 \@ifx{#1\undefined}
}%
\providecommand \@ifnum [1]{%
 \ifnum #1\expandafter \@firstoftwo
 \else \expandafter \@secondoftwo
 \fi
}%
\providecommand \@ifx [1]{%
 \ifx #1\expandafter \@firstoftwo
 \else \expandafter \@secondoftwo
 \fi
}%
\providecommand \natexlab [1]{#1}%
\providecommand \enquote  [1]{``#1''}%
\providecommand \bibnamefont  [1]{#1}%
\providecommand \bibfnamefont [1]{#1}%
\providecommand \citenamefont [1]{#1}%
\providecommand \href@noop [0]{\@secondoftwo}%
\providecommand \href [0]{\begingroup \@sanitize@url \@href}%
\providecommand \@href[1]{\@@startlink{#1}\@@href}%
\providecommand \@@href[1]{\endgroup#1\@@endlink}%
\providecommand \@sanitize@url [0]{\catcode `\\12\catcode `\$12\catcode
  `\&12\catcode `\#12\catcode `\^12\catcode `\_12\catcode `\%12\relax}%
\providecommand \@@startlink[1]{}%
\providecommand \@@endlink[0]{}%
\providecommand \url  [0]{\begingroup\@sanitize@url \@url }%
\providecommand \@url [1]{\endgroup\@href {#1}{\urlprefix }}%
\providecommand \urlprefix  [0]{URL }%
\providecommand \Eprint [0]{\href }%
\providecommand \doibase [0]{https://doi.org/}%
\providecommand \selectlanguage [0]{\@gobble}%
\providecommand \bibinfo  [0]{\@secondoftwo}%
\providecommand \bibfield  [0]{\@secondoftwo}%
\providecommand \translation [1]{[#1]}%
\providecommand \BibitemOpen [0]{}%
\providecommand \bibitemStop [0]{}%
\providecommand \bibitemNoStop [0]{.\EOS\space}%
\providecommand \EOS [0]{\spacefactor3000\relax}%
\providecommand \BibitemShut  [1]{\csname bibitem#1\endcsname}%
\let\auto@bib@innerbib\@empty
\bibitem [{\citenamefont {Thouless}(1983)}]{thouless1}%
  \BibitemOpen
  \bibfield  {author} {\bibinfo {author} {\bibfnamefont {D.~J.}\ \bibnamefont
  {Thouless}},\ }\bibfield  {title} {\bibinfo {title} {Quantization of particle
  transport},\ }\href {https://doi.org/10.1103/PhysRevB.27.6083} {\bibfield
  {journal} {\bibinfo  {journal} {Phys. Rev. B}\ }\textbf {\bibinfo {volume}
  {27}},\ \bibinfo {pages} {6083} (\bibinfo {year} {1983})}\BibitemShut
  {NoStop}%
\bibitem [{\citenamefont {Niu}\ and\ \citenamefont
  {Thouless}(1984)}]{thouless2}%
  \BibitemOpen
  \bibfield  {author} {\bibinfo {author} {\bibfnamefont {Q.}~\bibnamefont
  {Niu}}\ and\ \bibinfo {author} {\bibfnamefont {D.~J.}\ \bibnamefont
  {Thouless}},\ }\bibfield  {title} {\bibinfo {title} {Quantised adiabatic
  charge transport in the presence of substrate disorder and many-body
  interaction},\ }\href {https://doi.org/10.1088/0305-4470/17/12/016}
  {\bibfield  {journal} {\bibinfo  {journal} {Journal of Physics A:
  Mathematical and General}\ }\textbf {\bibinfo {volume} {17}},\ \bibinfo
  {pages} {2453} (\bibinfo {year} {1984})}\BibitemShut {NoStop}%
\bibitem [{\citenamefont {Bergholtz}\ \emph {et~al.}(2021)\citenamefont
  {Bergholtz}, \citenamefont {Budich},\ and\ \citenamefont
  {Kunst}}]{emil_review}%
  \BibitemOpen
  \bibfield  {author} {\bibinfo {author} {\bibfnamefont {E.~J.}\ \bibnamefont
  {Bergholtz}}, \bibinfo {author} {\bibfnamefont {J.~C.}\ \bibnamefont
  {Budich}},\ and\ \bibinfo {author} {\bibfnamefont {F.~K.}\ \bibnamefont
  {Kunst}},\ }\bibfield  {title} {\bibinfo {title} {Exceptional topology of
  non-{H}ermitian systems},\ }\href
  {https://doi.org/10.1103/RevModPhys.93.015005} {\bibfield  {journal}
  {\bibinfo  {journal} {Rev. Mod. Phys.}\ }\textbf {\bibinfo {volume} {93}},\
  \bibinfo {pages} {015005} (\bibinfo {year} {2021})}\BibitemShut {NoStop}%
\bibitem [{\citenamefont {Kunst}\ \emph
  {et~al.}(2018{\natexlab{a}})\citenamefont {Kunst}, \citenamefont
  {Edvardsson}, \citenamefont {Budich},\ and\ \citenamefont
  {Bergholtz}}]{flore-elisabet}%
  \BibitemOpen
  \bibfield  {author} {\bibinfo {author} {\bibfnamefont {F.~K.}\ \bibnamefont
  {Kunst}}, \bibinfo {author} {\bibfnamefont {E.}~\bibnamefont {Edvardsson}},
  \bibinfo {author} {\bibfnamefont {J.~C.}\ \bibnamefont {Budich}},\ and\
  \bibinfo {author} {\bibfnamefont {E.~J.}\ \bibnamefont {Bergholtz}},\
  }\bibfield  {title} {\bibinfo {title} {Biorthogonal bulk-boundary
  correspondence in non-{H}ermitian systems},\ }\href
  {https://doi.org/10.1103/PhysRevLett.121.026808} {\bibfield  {journal}
  {\bibinfo  {journal} {Phys. Rev. Lett.}\ }\textbf {\bibinfo {volume} {121}},\
  \bibinfo {pages} {026808} (\bibinfo {year} {2018}{\natexlab{a}})}\BibitemShut
  {NoStop}%
\bibitem [{\citenamefont {Edvardsson}\ \emph {et~al.}(2020)\citenamefont
  {Edvardsson}, \citenamefont {Kunst}, \citenamefont {Yoshida},\ and\
  \citenamefont {Bergholtz}}]{flore-elisabet0}%
  \BibitemOpen
  \bibfield  {author} {\bibinfo {author} {\bibfnamefont {E.}~\bibnamefont
  {Edvardsson}}, \bibinfo {author} {\bibfnamefont {F.~K.}\ \bibnamefont
  {Kunst}}, \bibinfo {author} {\bibfnamefont {T.}~\bibnamefont {Yoshida}},\
  and\ \bibinfo {author} {\bibfnamefont {E.~J.}\ \bibnamefont {Bergholtz}},\
  }\bibfield  {title} {\bibinfo {title} {{Phase transitions and generalized
  biorthogonal polarization in non-Hermitian systems}},\ }\href
  {https://doi.org/10.1103/PhysRevResearch.2.043046} {\bibfield  {journal}
  {\bibinfo  {journal} {Phys. Rev. Research}\ }\textbf {\bibinfo {volume}
  {2}},\ \bibinfo {pages} {043046} (\bibinfo {year} {2020})}\BibitemShut
  {NoStop}%
\bibitem [{\citenamefont {{Zelenayova}}\ and\ \citenamefont
  {{Bergholtz}}(2024)}]{maria_emil}%
  \BibitemOpen
  \bibfield  {author} {\bibinfo {author} {\bibfnamefont {M.}~\bibnamefont
  {{Zelenayova}}}\ and\ \bibinfo {author} {\bibfnamefont {E.~J.}\ \bibnamefont
  {{Bergholtz}}},\ }\bibfield  {title} {\bibinfo {title} {{Non-Hermitian
  extended midgap states and bound states in the continuum}},\ }\href
  {https://doi.org/10.1063/5.0184935} {\bibfield  {journal} {\bibinfo
  {journal} {Applied Physics Letters}\ }\textbf {\bibinfo {volume} {124}},\
  \bibinfo {eid} {041105} (\bibinfo {year} {2024})}\BibitemShut {NoStop}%
\bibitem [{\citenamefont {Mandal}(2024)}]{ips-bp}%
  \BibitemOpen
  \bibfield  {author} {\bibinfo {author} {\bibfnamefont {I.}~\bibnamefont
  {Mandal}},\ }\bibfield  {title} {\bibinfo {title} {{Identifying gap-closings
  in open non-Hermitian systems by biorthogonal polarization}},\ }\href
  {https://doi.org/10.1063/5.0198855} {\bibfield  {journal} {\bibinfo
  {journal} {Journal of Applied Physics}\ }\textbf {\bibinfo {volume} {135}},\
  \bibinfo {pages} {094402} (\bibinfo {year} {2024})}\BibitemShut {NoStop}%
\bibitem [{\citenamefont {Yang}\ \emph {et~al.}(2021)\citenamefont {Yang},
  \citenamefont {Morampudi},\ and\ \citenamefont {Bergholtz}}]{kang-emil}%
  \BibitemOpen
  \bibfield  {author} {\bibinfo {author} {\bibfnamefont {K.}~\bibnamefont
  {Yang}}, \bibinfo {author} {\bibfnamefont {S.~C.}\ \bibnamefont
  {Morampudi}},\ and\ \bibinfo {author} {\bibfnamefont {E.~J.}\ \bibnamefont
  {Bergholtz}},\ }\bibfield  {title} {\bibinfo {title} {Exceptional spin
  liquids from couplings to the environment},\ }\href
  {https://doi.org/10.1103/PhysRevLett.126.077201} {\bibfield  {journal}
  {\bibinfo  {journal} {Phys. Rev. Lett.}\ }\textbf {\bibinfo {volume} {126}},\
  \bibinfo {pages} {077201} (\bibinfo {year} {2021})}\BibitemShut {NoStop}%
\bibitem [{\citenamefont {{Mandal}}(2024)}]{ips-yao-lee}%
  \BibitemOpen
  \bibfield  {author} {\bibinfo {author} {\bibfnamefont {I.}~\bibnamefont
  {{Mandal}}},\ }\bibfield  {title} {\bibinfo {title} {{Non-Hermitian
  generalizations of the Yao-Lee model augmented by SO(3)-symmetry-breaking
  terms}},\ }\href {https://doi.org/10.1063/5.0209922} {\bibfield  {journal}
  {\bibinfo  {journal} {APL Quantum}\ }\textbf {\bibinfo {volume} {1}},\
  \bibinfo {pages} {036104} (\bibinfo {year} {2024})}\BibitemShut {NoStop}%
\bibitem [{\citenamefont {Chiu}\ \emph {et~al.}(2016)\citenamefont {Chiu},
  \citenamefont {Teo}, \citenamefont {Schnyder},\ and\ \citenamefont
  {Ryu}}]{schnyder_review}%
  \BibitemOpen
  \bibfield  {author} {\bibinfo {author} {\bibfnamefont {C.-K.}\ \bibnamefont
  {Chiu}}, \bibinfo {author} {\bibfnamefont {J.~C.~Y.}\ \bibnamefont {Teo}},
  \bibinfo {author} {\bibfnamefont {A.~P.}\ \bibnamefont {Schnyder}},\ and\
  \bibinfo {author} {\bibfnamefont {S.}~\bibnamefont {Ryu}},\ }\bibfield
  {title} {\bibinfo {title} {Classification of topological quantum matter with
  symmetries},\ }\href {https://doi.org/10.1103/RevModPhys.88.035005}
  {\bibfield  {journal} {\bibinfo  {journal} {Rev. Mod. Phys.}\ }\textbf
  {\bibinfo {volume} {88}},\ \bibinfo {pages} {035005} (\bibinfo {year}
  {2016})}\BibitemShut {NoStop}%
\bibitem [{\citenamefont {Bissonnette}\ \emph {et~al.}(2024)\citenamefont
  {Bissonnette}, \citenamefont {Delnour}, \citenamefont {Mckenna},
  \citenamefont {Eleuch}, \citenamefont {Hilke},\ and\ \citenamefont
  {MacKenzie}}]{ref_demand}%
  \BibitemOpen
  \bibfield  {author} {\bibinfo {author} {\bibfnamefont {A.}~\bibnamefont
  {Bissonnette}}, \bibinfo {author} {\bibfnamefont {N.}~\bibnamefont
  {Delnour}}, \bibinfo {author} {\bibfnamefont {A.}~\bibnamefont {Mckenna}},
  \bibinfo {author} {\bibfnamefont {H.}~\bibnamefont {Eleuch}}, \bibinfo
  {author} {\bibfnamefont {M.}~\bibnamefont {Hilke}},\ and\ \bibinfo {author}
  {\bibfnamefont {R.}~\bibnamefont {MacKenzie}},\ }\bibfield  {title} {\bibinfo
  {title} {{Boundary-induced topological transition in an open
  Su-Schrieffer-Heeger model}},\ }\href
  {https://doi.org/10.1103/PhysRevB.109.075106} {\bibfield  {journal} {\bibinfo
   {journal} {Phys. Rev. B}\ }\textbf {\bibinfo {volume} {109}},\ \bibinfo
  {pages} {075106} (\bibinfo {year} {2024})}\BibitemShut {NoStop}%
\bibitem [{\citenamefont {Niu}\ \emph {et~al.}(2012)\citenamefont {Niu},
  \citenamefont {Chung}, \citenamefont {Hsu}, \citenamefont {Mandal},
  \citenamefont {Raghu},\ and\ \citenamefont {Chakravarty}}]{ips-sudip}%
  \BibitemOpen
  \bibfield  {author} {\bibinfo {author} {\bibfnamefont {Y.}~\bibnamefont
  {Niu}}, \bibinfo {author} {\bibfnamefont {S.~B.}\ \bibnamefont {Chung}},
  \bibinfo {author} {\bibfnamefont {C.-H.}\ \bibnamefont {Hsu}}, \bibinfo
  {author} {\bibfnamefont {I.}~\bibnamefont {Mandal}}, \bibinfo {author}
  {\bibfnamefont {S.}~\bibnamefont {Raghu}},\ and\ \bibinfo {author}
  {\bibfnamefont {S.}~\bibnamefont {Chakravarty}},\ }\bibfield  {title}
  {\bibinfo {title} {{Majorana zero modes in a quantum Ising chain with
  longer-ranged interactions}},\ }\href
  {https://doi.org/10.1103/PhysRevB.85.035110} {\bibfield  {journal} {\bibinfo
  {journal} {Phys. Rev. B}\ }\textbf {\bibinfo {volume} {85}},\ \bibinfo
  {pages} {035110} (\bibinfo {year} {2012})}\BibitemShut {NoStop}%
\bibitem [{\citenamefont {Mandal}(2015)}]{ips-ep-epl}%
  \BibitemOpen
  \bibfield  {author} {\bibinfo {author} {\bibfnamefont {I.}~\bibnamefont
  {Mandal}},\ }\bibfield  {title} {\bibinfo {title} {{Exceptional points for
  chiral Majorana fermions in arbitrary dimensions}},\ }\href
  {https://doi.org/10.1209/0295-5075/110/67005} {\bibfield  {journal} {\bibinfo
   {journal} {EPL}\ }\textbf {\bibinfo {volume} {110}},\ \bibinfo {pages}
  {67005} (\bibinfo {year} {2015})}\BibitemShut {NoStop}%
\bibitem [{\citenamefont {Mandal}\ and\ \citenamefont
  {Tewari}(2016)}]{ips-tewari}%
  \BibitemOpen
  \bibfield  {author} {\bibinfo {author} {\bibfnamefont {I.}~\bibnamefont
  {Mandal}}\ and\ \bibinfo {author} {\bibfnamefont {S.}~\bibnamefont
  {Tewari}},\ }\bibfield  {title} {\bibinfo {title} {{Exceptional point
  description of one-dimensional chiral topological superconductors/superfluids
  in BDI class}},\ }\href {https://doi.org/10.1016/j.physe.2015.12.009}
  {\bibfield  {journal} {\bibinfo  {journal} {Physica E}\ }\textbf {\bibinfo
  {volume} {79}},\ \bibinfo {pages} {180} (\bibinfo {year} {2016})}\BibitemShut
  {NoStop}%
\bibitem [{\citenamefont {Su}\ \emph {et~al.}(1979)\citenamefont {Su},
  \citenamefont {Schrieffer},\ and\ \citenamefont {Heeger}}]{ssh0}%
  \BibitemOpen
  \bibfield  {author} {\bibinfo {author} {\bibfnamefont {W.~P.}\ \bibnamefont
  {Su}}, \bibinfo {author} {\bibfnamefont {J.~R.}\ \bibnamefont {Schrieffer}},\
  and\ \bibinfo {author} {\bibfnamefont {A.~J.}\ \bibnamefont {Heeger}},\
  }\bibfield  {title} {\bibinfo {title} {Solitons in polyacetylene},\ }\href
  {https://doi.org/10.1103/PhysRevLett.42.1698} {\bibfield  {journal} {\bibinfo
   {journal} {Phys. Rev. Lett.}\ }\textbf {\bibinfo {volume} {42}},\ \bibinfo
  {pages} {1698} (\bibinfo {year} {1979})}\BibitemShut {NoStop}%
\bibitem [{\citenamefont {Su}\ \emph {et~al.}(1980)\citenamefont {Su},
  \citenamefont {Schrieffer},\ and\ \citenamefont {Heeger}}]{ssh}%
  \BibitemOpen
  \bibfield  {author} {\bibinfo {author} {\bibfnamefont {W.~P.}\ \bibnamefont
  {Su}}, \bibinfo {author} {\bibfnamefont {J.~R.}\ \bibnamefont {Schrieffer}},\
  and\ \bibinfo {author} {\bibfnamefont {A.~J.}\ \bibnamefont {Heeger}},\
  }\bibfield  {title} {\bibinfo {title} {Soliton excitations in
  polyacetylene},\ }\href {https://doi.org/10.1103/PhysRevB.22.2099} {\bibfield
   {journal} {\bibinfo  {journal} {Phys. Rev. B}\ }\textbf {\bibinfo {volume}
  {22}},\ \bibinfo {pages} {2099} (\bibinfo {year} {1980})}\BibitemShut
  {NoStop}%
\bibitem [{\citenamefont {Yao}\ and\ \citenamefont {Wang}(2018)}]{yaowang}%
  \BibitemOpen
  \bibfield  {author} {\bibinfo {author} {\bibfnamefont {S.}~\bibnamefont
  {Yao}}\ and\ \bibinfo {author} {\bibfnamefont {Z.}~\bibnamefont {Wang}},\
  }\bibfield  {title} {\bibinfo {title} {Edge states and topological invariants
  of non-{H}ermitian systems},\ }\href
  {https://doi.org/10.1103/PhysRevLett.121.086803} {\bibfield  {journal}
  {\bibinfo  {journal} {Phys. Rev. Lett.}\ }\textbf {\bibinfo {volume} {121}},\
  \bibinfo {pages} {086803} (\bibinfo {year} {2018})}\BibitemShut {NoStop}%
\bibitem [{\citenamefont {Song}\ \emph {et~al.}(2019)\citenamefont {Song},
  \citenamefont {Yao},\ and\ \citenamefont {Wang}}]{song_yao_ssh}%
  \BibitemOpen
  \bibfield  {author} {\bibinfo {author} {\bibfnamefont {F.}~\bibnamefont
  {Song}}, \bibinfo {author} {\bibfnamefont {S.}~\bibnamefont {Yao}},\ and\
  \bibinfo {author} {\bibfnamefont {Z.}~\bibnamefont {Wang}},\ }\bibfield
  {title} {\bibinfo {title} {Non-{H}ermitian topological invariants in real
  space},\ }\href {https://doi.org/10.1103/PhysRevLett.123.246801} {\bibfield
  {journal} {\bibinfo  {journal} {Phys. Rev. Lett.}\ }\textbf {\bibinfo
  {volume} {123}},\ \bibinfo {pages} {246801} (\bibinfo {year}
  {2019})}\BibitemShut {NoStop}%
\bibitem [{\citenamefont {{Han}}\ \emph {et~al.}(2021)\citenamefont {{Han}},
  \citenamefont {{Liu}},\ and\ \citenamefont {{Liu}}}]{han_ssh}%
  \BibitemOpen
  \bibfield  {author} {\bibinfo {author} {\bibfnamefont {Y.~Z.}\ \bibnamefont
  {{Han}}}, \bibinfo {author} {\bibfnamefont {J.~S.}\ \bibnamefont {{Liu}}},\
  and\ \bibinfo {author} {\bibfnamefont {C.~S.}\ \bibnamefont {{Liu}}},\
  }\bibfield  {title} {\bibinfo {title} {{The topological counterparts of
  non-Hermitian SSH models}},\ }\href
  {https://doi.org/10.1088/1367-2630/ac3e9f} {\bibfield  {journal} {\bibinfo
  {journal} {New Journal of Physics}\ }\textbf {\bibinfo {volume} {23}},\
  \bibinfo {eid} {123029} (\bibinfo {year} {2021})}\BibitemShut {NoStop}%
\bibitem [{\citenamefont {Lee}(2016)}]{ssh_lee}%
  \BibitemOpen
  \bibfield  {author} {\bibinfo {author} {\bibfnamefont {T.~E.}\ \bibnamefont
  {Lee}},\ }\bibfield  {title} {\bibinfo {title} {Anomalous edge state in a
  non-{H}ermitian lattice},\ }\href
  {https://doi.org/10.1103/PhysRevLett.116.133903} {\bibfield  {journal}
  {\bibinfo  {journal} {Phys. Rev. Lett.}\ }\textbf {\bibinfo {volume} {116}},\
  \bibinfo {pages} {133903} (\bibinfo {year} {2016})}\BibitemShut {NoStop}%
\bibitem [{\citenamefont {Yin}\ \emph {et~al.}(2018)\citenamefont {Yin},
  \citenamefont {Jiang}, \citenamefont {Li}, \citenamefont {L\"u},\ and\
  \citenamefont {Chen}}]{yinjiangliluchen}%
  \BibitemOpen
  \bibfield  {author} {\bibinfo {author} {\bibfnamefont {C.}~\bibnamefont
  {Yin}}, \bibinfo {author} {\bibfnamefont {H.}~\bibnamefont {Jiang}}, \bibinfo
  {author} {\bibfnamefont {L.}~\bibnamefont {Li}}, \bibinfo {author}
  {\bibfnamefont {R.}~\bibnamefont {L\"u}},\ and\ \bibinfo {author}
  {\bibfnamefont {S.}~\bibnamefont {Chen}},\ }\bibfield  {title} {\bibinfo
  {title} {Geometrical meaning of winding number and its characterization of
  topological phases in one-dimensional chiral non-{H}ermitian systems},\
  }\href {https://doi.org/10.1103/PhysRevA.97.052115} {\bibfield  {journal}
  {\bibinfo  {journal} {Phys. Rev. A}\ }\textbf {\bibinfo {volume} {97}},\
  \bibinfo {pages} {052115} (\bibinfo {year} {2018})}\BibitemShut {NoStop}%
\bibitem [{\citenamefont {Rice}\ and\ \citenamefont {Mele}(1982)}]{ricemele}%
  \BibitemOpen
  \bibfield  {author} {\bibinfo {author} {\bibfnamefont {M.~J.}\ \bibnamefont
  {Rice}}\ and\ \bibinfo {author} {\bibfnamefont {E.~J.}\ \bibnamefont
  {Mele}},\ }\bibfield  {title} {\bibinfo {title} {Elementary excitations of a
  linearly conjugated diatomic polymer},\ }\href
  {https://doi.org/10.1103/PhysRevLett.49.1455} {\bibfield  {journal} {\bibinfo
   {journal} {Phys. Rev. Lett.}\ }\textbf {\bibinfo {volume} {49}},\ \bibinfo
  {pages} {1455} (\bibinfo {year} {1982})}\BibitemShut {NoStop}%
\bibitem [{\citenamefont {{Atala}}\ \emph {et~al.}(2013)\citenamefont
  {{Atala}}, \citenamefont {{Aidelsburger}}, \citenamefont {{Barreiro}},
  \citenamefont {{Abanin}}, \citenamefont {{Kitagawa}}, \citenamefont
  {{Demler}},\ and\ \citenamefont {{Bloch}}}]{ssh_expt}%
  \BibitemOpen
  \bibfield  {author} {\bibinfo {author} {\bibfnamefont {M.}~\bibnamefont
  {{Atala}}}, \bibinfo {author} {\bibfnamefont {M.}~\bibnamefont
  {{Aidelsburger}}}, \bibinfo {author} {\bibfnamefont {J.~T.}\ \bibnamefont
  {{Barreiro}}}, \bibinfo {author} {\bibfnamefont {D.}~\bibnamefont
  {{Abanin}}}, \bibinfo {author} {\bibfnamefont {T.}~\bibnamefont
  {{Kitagawa}}}, \bibinfo {author} {\bibfnamefont {E.}~\bibnamefont
  {{Demler}}},\ and\ \bibinfo {author} {\bibfnamefont {I.}~\bibnamefont
  {{Bloch}}},\ }\bibfield  {title} {\bibinfo {title} {{Direct measurement of
  the Zak phase in topological Bloch bands}},\ }\href
  {https://doi.org/10.1038/nphys2790} {\bibfield  {journal} {\bibinfo
  {journal} {Nature Physics}\ }\textbf {\bibinfo {volume} {9}},\ \bibinfo
  {pages} {795} (\bibinfo {year} {2013})}\BibitemShut {NoStop}%
\bibitem [{\citenamefont {{Drost}}\ \emph {et~al.}(2017)\citenamefont
  {{Drost}}, \citenamefont {{Ojanen}}, \citenamefont {{Harju}},\ and\
  \citenamefont {{Liljeroth}}}]{ssh_expt2}%
  \BibitemOpen
  \bibfield  {author} {\bibinfo {author} {\bibfnamefont {R.}~\bibnamefont
  {{Drost}}}, \bibinfo {author} {\bibfnamefont {T.}~\bibnamefont {{Ojanen}}},
  \bibinfo {author} {\bibfnamefont {A.}~\bibnamefont {{Harju}}},\ and\ \bibinfo
  {author} {\bibfnamefont {P.}~\bibnamefont {{Liljeroth}}},\ }\bibfield
  {title} {\bibinfo {title} {{Topological states in engineered atomic
  lattices}},\ }\href {https://doi.org/10.1038/nphys4080} {\bibfield  {journal}
  {\bibinfo  {journal} {Nature Physics}\ }\textbf {\bibinfo {volume} {13}},\
  \bibinfo {pages} {668} (\bibinfo {year} {2017})}\BibitemShut {NoStop}%
\bibitem [{\citenamefont {Przysiezna}\ \emph {et~al.}(2015)\citenamefont
  {Przysiezna}, \citenamefont {Dutta},\ and\ \citenamefont
  {Zakrzewski}}]{ricemele_optical}%
  \BibitemOpen
  \bibfield  {author} {\bibinfo {author} {\bibfnamefont {A.}~\bibnamefont
  {Przysiezna}}, \bibinfo {author} {\bibfnamefont {O.}~\bibnamefont {Dutta}},\
  and\ \bibinfo {author} {\bibfnamefont {J.}~\bibnamefont {Zakrzewski}},\
  }\bibfield  {title} {\bibinfo {title} {{Rice-Mele model with topological
  solitons in an optical lattice}},\ }\href
  {https://doi.org/10.1088/1367-2630/17/1/013018} {\bibfield  {journal}
  {\bibinfo  {journal} {New J. Phys.}\ }\textbf {\bibinfo {volume} {17}},\
  \bibinfo {pages} {013018} (\bibinfo {year} {2015})}\BibitemShut {NoStop}%
\bibitem [{\citenamefont {{Brody}}(2014)}]{brody}%
  \BibitemOpen
  \bibfield  {author} {\bibinfo {author} {\bibfnamefont {D.~C.}\ \bibnamefont
  {{Brody}}},\ }\bibfield  {title} {\bibinfo {title} {{Biorthogonal quantum
  mechanics}},\ }\href {https://doi.org/10.1088/1751-8113/47/3/035305}
  {\bibfield  {journal} {\bibinfo  {journal} {Journal of Physics A Mathematical
  General}\ }\textbf {\bibinfo {volume} {47}},\ \bibinfo {eid} {035305}
  (\bibinfo {year} {2014})}\BibitemShut {NoStop}%
\bibitem [{\citenamefont {Zhang}\ \emph {et~al.}(2024)\citenamefont {Zhang},
  \citenamefont {Li}, \citenamefont {Luo},\ and\ \citenamefont
  {Yi}}]{zhenming}%
  \BibitemOpen
  \bibfield  {author} {\bibinfo {author} {\bibfnamefont {Z.}~\bibnamefont
  {Zhang}}, \bibinfo {author} {\bibfnamefont {T.}~\bibnamefont {Li}}, \bibinfo
  {author} {\bibfnamefont {X.-W.}\ \bibnamefont {Luo}},\ and\ \bibinfo {author}
  {\bibfnamefont {W.}~\bibnamefont {Yi}},\ }\bibfield  {title} {\bibinfo
  {title} {{Biorthogonal topological charge pumping in non-Hermitian
  systems}},\ }\href {https://doi.org/10.1103/PhysRevB.109.224307} {\bibfield
  {journal} {\bibinfo  {journal} {Phys. Rev. B}\ }\textbf {\bibinfo {volume}
  {109}},\ \bibinfo {pages} {224307} (\bibinfo {year} {2024})}\BibitemShut
  {NoStop}%
\bibitem [{\citenamefont {Ib\'a\~nez}\ and\ \citenamefont {Muga}(2014)}]{muga}%
  \BibitemOpen
  \bibfield  {author} {\bibinfo {author} {\bibfnamefont {S.}~\bibnamefont
  {Ib\'a\~nez}}\ and\ \bibinfo {author} {\bibfnamefont {J.~G.}\ \bibnamefont
  {Muga}},\ }\bibfield  {title} {\bibinfo {title} {{Adiabaticity condition for
  non-Hermitian Hamiltonians}},\ }\href
  {https://doi.org/10.1103/PhysRevA.89.033403} {\bibfield  {journal} {\bibinfo
  {journal} {Phys. Rev. A}\ }\textbf {\bibinfo {volume} {89}},\ \bibinfo
  {pages} {033403} (\bibinfo {year} {2014})}\BibitemShut {NoStop}%
\bibitem [{\citenamefont {Kunst}\ \emph
  {et~al.}(2018{\natexlab{b}})\citenamefont {Kunst}, \citenamefont {van
  Miert},\ and\ \citenamefont {Bergholtz}}]{flore_2d_chern}%
  \BibitemOpen
  \bibfield  {author} {\bibinfo {author} {\bibfnamefont {F.~K.}\ \bibnamefont
  {Kunst}}, \bibinfo {author} {\bibfnamefont {G.}~\bibnamefont {van Miert}},\
  and\ \bibinfo {author} {\bibfnamefont {E.~J.}\ \bibnamefont {Bergholtz}},\
  }\bibfield  {title} {\bibinfo {title} {Lattice models with exactly solvable
  topological hinge and corner states},\ }\href
  {https://doi.org/10.1103/PhysRevB.97.241405} {\bibfield  {journal} {\bibinfo
  {journal} {Phys. Rev. B}\ }\textbf {\bibinfo {volume} {97}},\ \bibinfo
  {pages} {241405} (\bibinfo {year} {2018}{\natexlab{b}})}\BibitemShut
  {NoStop}%
\bibitem [{\citenamefont {Yokomizo}\ and\ \citenamefont
  {Murakami}(2019)}]{yokomizo}%
  \BibitemOpen
  \bibfield  {author} {\bibinfo {author} {\bibfnamefont {K.}~\bibnamefont
  {Yokomizo}}\ and\ \bibinfo {author} {\bibfnamefont {S.}~\bibnamefont
  {Murakami}},\ }\bibfield  {title} {\bibinfo {title} {Non-bloch band theory of
  non-{H}ermitian systems},\ }\href
  {https://doi.org/10.1103/PhysRevLett.123.066404} {\bibfield  {journal}
  {\bibinfo  {journal} {Phys. Rev. Lett.}\ }\textbf {\bibinfo {volume} {123}},\
  \bibinfo {pages} {066404} (\bibinfo {year} {2019})}\BibitemShut {NoStop}%
\bibitem [{\citenamefont {Mandal}\ and\ \citenamefont
  {Bergholtz}(2021)}]{ips-emil}%
  \BibitemOpen
  \bibfield  {author} {\bibinfo {author} {\bibfnamefont {I.}~\bibnamefont
  {Mandal}}\ and\ \bibinfo {author} {\bibfnamefont {E.~J.}\ \bibnamefont
  {Bergholtz}},\ }\bibfield  {title} {\bibinfo {title} {Symmetry and
  higher-order exceptional points},\ }\href
  {https://doi.org/10.1103/PhysRevLett.127.186601} {\bibfield  {journal}
  {\bibinfo  {journal} {Phys. Rev. Lett.}\ }\textbf {\bibinfo {volume} {127}},\
  \bibinfo {pages} {186601} (\bibinfo {year} {2021})}\BibitemShut {NoStop}%
\bibitem [{\citenamefont {Yang}\ and\ \citenamefont {Mandal}(2023)}]{ips-kang}%
  \BibitemOpen
  \bibfield  {author} {\bibinfo {author} {\bibfnamefont {K.}~\bibnamefont
  {Yang}}\ and\ \bibinfo {author} {\bibfnamefont {I.}~\bibnamefont {Mandal}},\
  }\bibfield  {title} {\bibinfo {title} {Enhanced eigenvector sensitivity and
  algebraic classification of sublattice-symmetric exceptional points},\ }\href
  {https://doi.org/10.1103/PhysRevB.107.144304} {\bibfield  {journal} {\bibinfo
   {journal} {Phys. Rev. B}\ }\textbf {\bibinfo {volume} {107}},\ \bibinfo
  {pages} {144304} (\bibinfo {year} {2023})}\BibitemShut {NoStop}%
\end{thebibliography}%
\end{document}